\documentclass[usenatbib]{mn2e}
\usepackage{epsfig}
\usepackage{amsmath,amssymb}

\usepackage[dvips]{color}

\newcommand{\Caf}{SDSS J102915+172927}
\newcommand{\percc}{{\rm cm^{-3}}}
\newcommand{\um}{{\rm \mu m}}
\newcommand{\E}[1]{\times 10^{#1}}

\newcommand{\Mg}{{\rm Mg}}

\newcommand{\Si}{{\rm Si}}
\newcommand{\Fe}{{\rm Fe}}

\newcommand{\norev}{{\tt norev}}
\newcommand{\revone}{{\tt rev1}}
\newcommand{\revtwo}{{\tt rev2}}
\newcommand{\revthree}{{\tt rev3}}

\newcommand{\Pyroxene}{{\rm MgSiO_3}}
\newcommand{\Olivine}{{\rm Mg_2SiO_4}}
\newcommand{\Magnetite}{{\rm Fe_3O_4}}
\newcommand{\Silica}{{\rm SiO_2}}

\newcommand{\Alumina}{{\rm Al_2O_3}}

\newcommand{\nH}{n_{{\rm H}}}
\newcommand{\mH}{m_{{\rm H}}}
\newcommand{\ngrowi}{n_{{\rm grow},i}}

\newcommand{\XH}{X_{{\rm H}}}
\newcommand{\Zsun}{Z_{\odot}}
\newcommand{\Msun}{M_{\odot}}

\newcommand{\Mpr}{M_{{\rm pr}}}

\newcommand{\fdepini}{f_{{\rm dep},0}}

\title[grain growth and the origin of SDSS J102915]{Dust grain growth and the formation of the extremely primitive star \Caf}

\author[G. Chiaki et al.]
{Gen Chiaki,$^{1}$\thanks{E-mail: gen.chiaki@utap.phys.s.u-tokyo.ac.jp}
Raffaella Schneider,$^{2}$
Takaya Nozawa,$^{3}$ 
Kazuyuki Omukai,$^{4}$
\newauthor 
Marco Limongi,$^{2,3}$
Naoki Yoshida,$^{1,3}$
Alessandro Chieffi$^{5}$ 
\\
$^{1}$Department of Physics, Graduate School of Science, The University of Tokyo, 
7-3-1 Hongo, Bunkyo, Tokyo 113-0033, Japan \\
$^{2}$INAF/Osservatorio Astronomico di Roma, via Frascati 33, 00040 Roma, Italy \\
$^{3}$Kavli Institute for the Physics and Mathematics of the Universe (WPI), 
Todai Institutes for Advanced Study, \\
The University of Tokyo, Kashiwa, Chiba 277-8583, Japan \\
$^{4}$Astronomical Institute, Tohoku University, Sendai 980-8578, Japan \\
$^{5}$INAF/IASF, Via Fosso del Cavaliere 100, 00133 Roma, Italy}
\begin{document}

\date{}

\pagerange{\pageref{firstpage}--\pageref{lastpage}} \pubyear{2013}

\maketitle

\label{firstpage}

\begin{abstract}
Dust grains in low-metallicity star-forming regions may be responsible for the formation
of the first low-mass stars. 
The minimal conditions to activate dust-induced fragmentation require
the gas to be pre-enriched above a critical dust-to-gas mass ratio 
${\cal D}_{\rm cr} = [2.6$--$6.3] \times 10^{-9}$ with the spread reflecting the dependence 
on the grain properties. 
The recently discovered Galactic halo star \Caf \ has a stellar mass of $0.8 \ \Msun$ and a metallicity
of $Z \sim 4.5 \times 10^{-5} \ \Zsun$ and represents an optimal candidate for the 
dust-induced low-mass star formation. Indeed, for the two most plausible Population III 
supernova progenitors, with $20 \ \Msun$ and $35 \ \Msun$, the critical dust-to-gas mass ratio can
be overcome provided that at least $0.4 \ \Msun$  of dust condenses in the ejecta, 
allowing for moderate destruction by the reverse shock. 
Here we show that even if dust formation in the first supernovae is less efficient or strong dust destruction 
does occur, grain growth during the collapse of the parent gas cloud is sufficiently rapid
to activate dust cooling and likely fragmentation into low-mass and long-lived stars. We find that the size distribution
of carbon grains is not modified by grain growth 
because at small densities,  below $\nH \sim 10^6 \ \percc$,  carbon atoms have been locked into CO
molecules. Silicates and magnetite grains can experience significant grain growth in the density range 
$10^9 \ \percc < \nH < 10^{12} \ \percc$ by accreting gas-phase species (SiO, SiO$_2$, and Fe) until their
gas-phase abundance drops to zero, reaching condensation efficiencies $\approx 1$.  
The corresponding increase in the dust-to-gas mass ratio allows dust-induced cooling and fragmentation to 
be activated at $10^{12} \ \percc < \nH < 10^{14} \ \percc$, 
before the collapsing cloud becomes optically thick to continuum radiation. 
We show that for all the initial conditions that apply to the parent cloud of \Caf, dust-driven
fragmentation is able to account for the formation of the star. 
We discuss the implication of this finding for the formation of the first low-mass stars in the 
Universe and derive some minimum critical abundances of refractory elements by our results:
$\rm [Mg/H] _{\rm cr} = [-5.02 : -4.73]$, 
$\rm [Si/H] _{\rm cr} = [-5.24  : -4.73]$, and 
$\rm [Fe/H] _{\rm cr} = [-5.43 : -3.70]$.
\end{abstract}



\section{INTRODUCTION}

The physical conditions which enable the first
low-mass and long-lived stars to form in the Universe is still
a subject of debate \citep{Karlsson13}.
Recent numerical simulations
of primordial star formation suggest that
the first stars could have a wide variety of masses 
\citep[][for a recent comprehensive review]{Greif12,Hirano13,Bromm13}.
Indeed, no evidence for the existence of primordial low-mass
stars has been found in surveys of the Galactic Halo and in nearby dwarfs
\citep{Beers05}.   

Once the first supernovae explode and start to seed the gas 
with metals and dust grains, the physical properties of 
star-forming regions change
\citep{Bromm03,Chiaki13SN}.
Low-metallicity
gas can achieve larger cooling rates through additional molecular 
species (OH, CO, H$_2$O), fine structure line-cooling (mostly O {\sc i}
and C {\sc ii}), and thermal emission from dust grains \citep{Omukai00,Bromm01,Schneider02,Omukai05}.
The relative importance of these coolants depends on the density (time)
regime during the collapse and on the initial metallicity and dust 
content of the collapsing core.   

O {\sc i} and C {\sc ii} fine-structure line cooling dominates the thermal evolution at
$\nH < 10^{4} \ \percc$ until the NLTE-LTE transition occurs
and the cooling efficiency decreases. 
\citet{BrommLoeb03} and \citet{Frebel07} 
have discussed the critical O and C abundances required to overcome
the compressional heating rate and fragment the gas, $\rm [C/H]_{\rm cr} = -3.5$
and $\rm [O/H]_{\rm cr} = -3.05$.
Given the typical gas temperatures and
densities where line-induced fragmentation takes place, the associated Jeans-unstable
fragment masses are still relatively large, $\ge 10 \ \Msun$ 
\citep{Schneider06,Safranek13}.

A pervasive fragmentation mode that allows the formation of solar or sub-solar mass 
fragments is activated at higher gas densities $\nH \sim 10^{12}$--$10^{14} \ \percc$ where
dust grains are collisionally excited and emit continuum radiation, thereby decreasing
the gas temperature, $T$, until it becomes coupled with the dust temperature,
or the gas becomes optically thick \citep{Schneider02}.
The efficiency of this physical process depends on the dust-to-gas mass ratio and on the 
total cross-section of the grains, with smaller grains providing a larger 
contribution to cooling \citep{Schneider06}. Hence, the minimal conditions for 
dust-induced fragmentation have been expressed in terms of a minimal critical dust-to-gas
mass ratio, 
\begin{equation}
S {\cal D_{\rm cr}} = 1.4\E{-3} \ {\rm cm^2 \ g^{-1}} \left( \frac{T}{10^3 \ {\rm K}} \right) ^{-1/2}
\left( \frac{\nH}{10^{12} \ \percc} \right)^{-1/2},
\label{eq:SD_crit}
\end{equation}
where $S$ is the total cross-section of dust grains per unit dust mass and the expression 
holds in the regime where dust cooling is effective, hence grain temperatures
are much lower than gas temperature. 
An extensive parameter-space exploration shows that ${\cal D_{\rm cr}} \sim  4.4 \times 10^{-9}$
can be considered as a good representative value for the minimal dust enrichment required
to activate dust-induced fragmentation \citep{Schneider12Crit}.

The recently discovered Galactic halo star \Caf \ represents a perfect candidate for dust-induced 
fragmentation \citep{Schneider12Caf, Klessen12}.
It has a stellar mass of $< 0.8 \ \Msun$ and a metallicity of only 
$Z_{\rm obs} = 4.5 \times 10^{-5} \ \Zsun$ being the most chemically pristine star ever observed in 
the Universe \citep{Caffau11Nat}. 
\citet{Schneider12Caf} investigate the origin of \Caf \ by
reconstructing the physical conditions of the parent cloud out of which the star had formed.
They assume pre-enrichment of the gas with metals and dust released by primordial core-collapse
supernovae, allowing for the partial destruction of grains by the reverse shock.
To simulate the physical conditions of the parent cloud of \Caf \ at the onset
of collapse, the resulting metal and dust yields are characterized by the initial 
depletion factor $\fdepini$,  defined to be the ratio of dust mass relative to the 
total mass of heavy elements
and are rescaled to the total metallicity of \Caf \, yielding an initial 
dust-to-gas mass ratio of ${\cal D}_0 = \fdepini Z_{\rm obs}$. 
Given these initial conditions, \citet{Schneider12Caf} follow the thermal evolution of the collapsing clouds
and find dust-induced fragmentation to be effective only in 4 out of the 8 models considered. 
Not surprisingly, since both $\fdepini$ and 
${\cal D}_0$ are assumed to be constant during the course of the collapse, the successful models conform
to the condition that $\fdepini > 0.01$, that is ${\cal D}_0 > {\cal D}_{\rm cr}$.  

\begin{figure*}
\includegraphics[width=15cm]{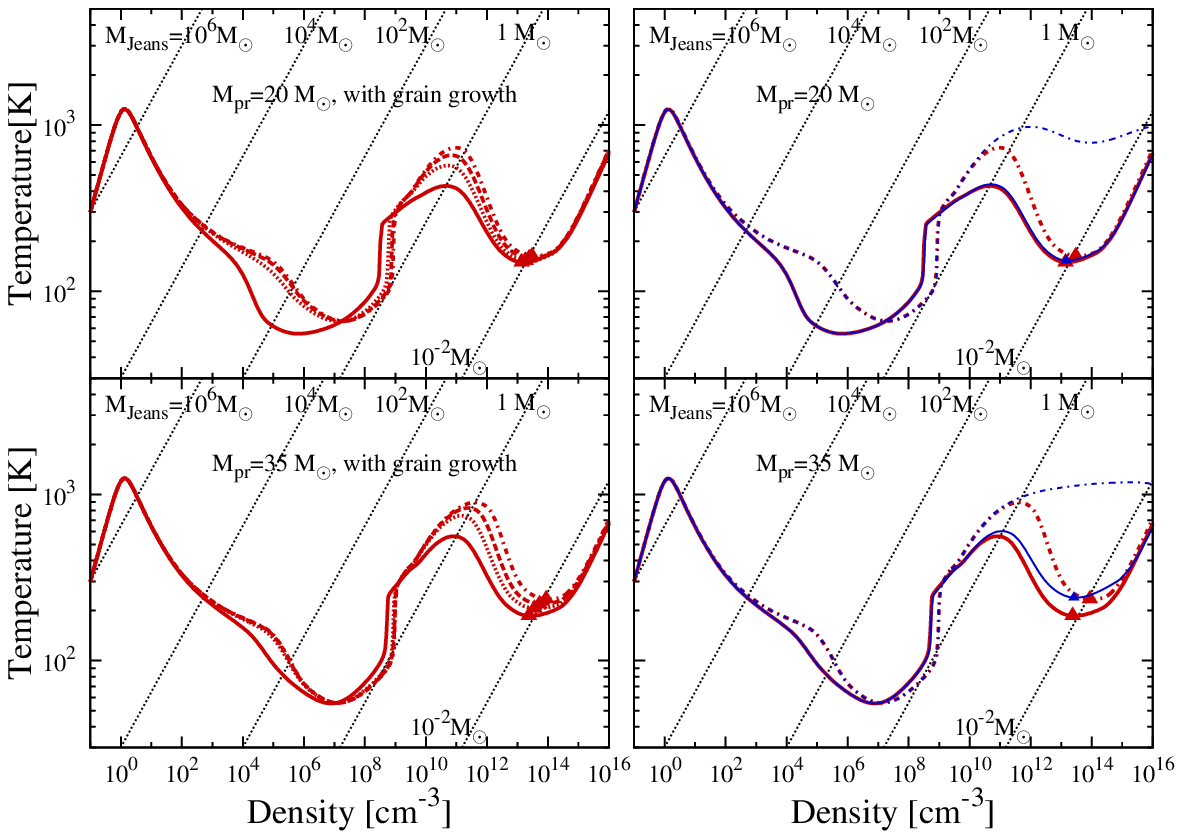}
\caption{
{\it Left panels:} the temperature evolution for the 
\norev \ (solid), \revone \ (dotted), \revtwo \ (dashed), and \revthree \ (dot-dashed) models with grain growth
for progenitor mass $\Mpr = 20 \ \Msun $ (top) and $35 \ \Msun$ (bottom) as a function of the 
cloud density. {\it Right panels:} comparison between the 
temperature evolution with (red) and without (blue) grain growth.
We plot the results for \norev \ (solid) and \revthree \ (dot-dashed) models.
Triangles identify the states where the fragmentation conditions are met.}
\label{fig:nT}
\end{figure*}

In a recent paper, \citet{Nozawa12} show that as the density in the collapsing cloud increases, 
dust grains can grow in mass by accreting metal species. Grain growth
during the collapse is very efficient and the dust-to-gas mass ratio increases 
even for very small initial gas metallicity and depletion factor. 
Earlier in \citet{Chiaki13single}, we have studied
the impact of grain growth on the thermal evolution of collapsing clouds. 
It is shown that even for small values of the initial
depletion factor, $\fdepini \approx  10^{-3}$, and for initial 
metallicities in the range $10^{-5.5}$--$10^{-4.5} \ \Zsun$, grain growth
enables to activate dust cooling at densities $\nH \sim 10^{12}$--$10^{14} \ \percc$,
thereby inducing fragmentation into low-mass clumps. However, the model is based
on a few simplifying assumptions, such as the use of 
an average grain size as a proxy to the full grain size distribution
and of a single dust species to dominate the total dust mass.  

The aim of the present study is to investigate the implications that the process
of grain growth may have on the formation of a low-mass star with a metallicity of 
$Z = 4.5 \E{-5} \ \Zsun$ such as \Caf. Starting from the same supernova progenitor 
models and initial conditions as \citet{Schneider12Caf}, we take into account the
modifications induced by grain growth on the size distribution of the grains
to check whether conditions favorable to fragmentation can be met.
A more systematic investigation of the parameter space, including different supernova
models and initial gas metallicities, will be presented in a forthcoming paper \citep{Chiaki13multi}.

\section{METHOD}

We base our calculation on a one-zone semi-analytical model that enables to 
follow the thermal evolution of the collapsing cloud and the process of 
grain growth in a self-consistent way. 
Here we describe 
the new features of the model required to implement
the process of grain growth. 
For each dust species initially present in the collapsing cloud,
we calculate the time evolution of condensation efficiency, $f_{ij}$,
defined as the {\it number} fraction of nuclei of 
element $j$ condensed into grains of species $i$.
We assume that ($i$) grains are
spherical particles, and that ($ii$) the rate is controlled by a 
single chemical species, referred to as the key species, which corresponds to
the reactant with the least collisional frequency on the target. 

Hence, the growth rate of a dust species $i$ can be expressed as, 
\begin{equation}
\left( \frac{dr}{dt} \right)_i = \alpha _i \left( \frac{4\pi}{3} a_{i,0}^3 \right)
\left( \frac{kT}{2\pi m_{i1}} \right)^{1/2} n_{i1} (t),
\label{GG:drdt}
\end{equation}
where $\alpha_i$ is the sticking probability of a gas particle onto grains of species $i$, 
$a_{i,0}$ is the radius of a single monomer of species $i$ in the condensed phase, 
and $m_{i1}$ and $n_{i1} $ are the mass 
and number density of the key species for grains of species $i$.\footnote{
The subscript $i1$ denotes the key {\it chemical species} for grains of species $i$
(e.g., Mg atoms or SiO moledules for $\Olivine$ grains).
We use a different subscript $j$ to indicate {\it elements}
(e.g., Mg, Si, and so on).}
Note that the growth rate is independent of the grain radius.
In this paper, we consider size distribution function $\varphi _{i} (r) $ 
for grains of species $i$.
The function is normalized as $\int \varphi _{i} (r) dr = 1$, and thus
the number density of dust $i$ with radii between $r$ and $r+dr$ is $n_i \varphi _i(r)dr$,
where $n_i$ is the total number density of dust grains $i$.
Given the size distribution at the time $t$, 
we can calculate the condensation efficiency of grains.
Condensation efficiency is obtained by integrating the size distribution as
\begin{equation}
f_{ij}(t) = f_{ij,0} 
\frac{\int  r^3 \varphi _i (r) dr}{\int r^3 \varphi _{i,0} (r) dr},
\label{GG:f}
\end{equation}
where the subscript `0' indicates the initial values.

As the collapse proceeds, equation (\ref{GG:drdt}) allows to compute the time-dependent grain
size distribution, dust mass and depletion factor. 
In addition, we need to compute the condensation efficiencies $f_{ij}$, defined as the number fraction of element $j$ condensed 
into dust species $i$.
We solve the non-equilibrium chemistry of eight primordial elements
H$^+$, $e^-$, H, H$^-$, H$_2$, D$^+$, D, and HD,
and nineteen heavy elements C$^+$, C, CH, CH$_2$, CO$^+$, CO, CO$_2$, O$^+$, O, OH$^+$, OH, 
H$_2$O$^+$, H$_2$O, H$_3$O$^+$, O$_2^+$, O$_2$, Si, SiO, and SiO$_2$.
Silicon atoms and SiO molecules are oxidized mainly by OH molecules
\citep{Langer90}.

We adopt the same initial conditions as \cite{Schneider12Caf} to understand how grain growth
may affect the resulting cloud fragmentation properties. 
In these models, dust species included are
forsterite ($\Olivine$), enstatite ($\Pyroxene$), magnetite ($\Magnetite$), 
amorphous carbon (C), silica ($\Silica$), and alumina ($\Alumina$).
Then, we assume that silicate ($\Olivine$ and $\Pyroxene$) and silica grains grow when 
SiO and SiO$_2$ molecules stick to the grains, respectively.
Dust temperature, dust continuum opacity and cooling rate are calculated at
each time using the time-dependent grain size distribution function on the basis of formulae
of \citet{Schneider06}.
The size bins have an initial width of 0.1 dex. 
The results are not significantly different from the ones obtained when
the initial size bins are 0.01 dex width.

Additional features of the model, other than the part of grain growth, are 
the same as in the model used by \citet{Schneider12Caf}.
The collapsing gas cloud has a 
total metallicity of $Z_{\rm obs} = 4.5 \times 10^{-5} Z_{\odot}$, equal to the observed metallicity of
\Caf. The initial metal and dust contents are taken from two different Population III supernova
models with progenitor masses and explosion energies selected to minimize the
scatter between the theoretical yields and the observed ones. Despite its very low iron content, 
SDSS J102915+172927 shows an abundance pattern consistent with typical Galactic halo signatures and
the best agreement is found for $20 \ \Msun $ and $35 \ \Msun$ progenitors with mass cuts, 
explosion energies and ejected $^{56}$Ni mass in the range of what is typically found for ordinary
core-collapse supernovae \citep{LimongiChieffi12}. Using these supernova models as input for the
dust nucleation model \citep{Bianchi07}, they compute the mass, composition and size distribution
of dust grains released in the interstellar medium, after allowing for the partial destruction of the newly formed
dust by the reverse shock. 
As is the same as \cite{Schneider12Caf}, for each progenitor mass we consider a model where no
dust destruction is assumed to take place (\norev) and three additional models with increasing
dust destruction efficiency by the reverse shock (\revone, \revtwo, and \revthree \ models which correspond
to circumstellar medium gas densities of $10^{-25}$, $10^{-24}$, and $10^{-23} \ {\rm g \ cm^{-3}}$, respectively).
The initial depletion factors for each of these eight models, $\fdepini$, 
are 0.26, 0.045, 0.015, 0.0057 for \norev, \revone, \revtwo, and \revthree \ models with $\Mpr = 20 \ \Msun$, respectively,
and 0.076, 0.0082, 0.0025, 0.0010 for \norev, \revone, \revtwo, and \revthree \ models with $\Mpr = 35 \ \Msun$, respectively.


\begin{figure}
\includegraphics[width=9cm]{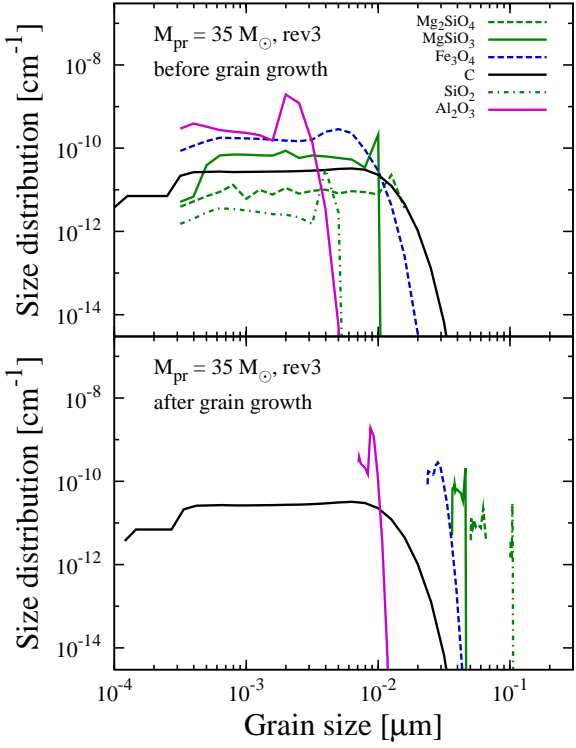}
\caption{
Size distribution functions, $n_i \varphi _i(r) / \nH$ of
different dust species before (top) and after (bottom) grain growth
for the \revthree \ model with $\Mpr = 35 \ \Msun$.
}
\label{fig:dist_S0_M35_rev3}
\end{figure}

\begin{figure}
\includegraphics[width=9cm]{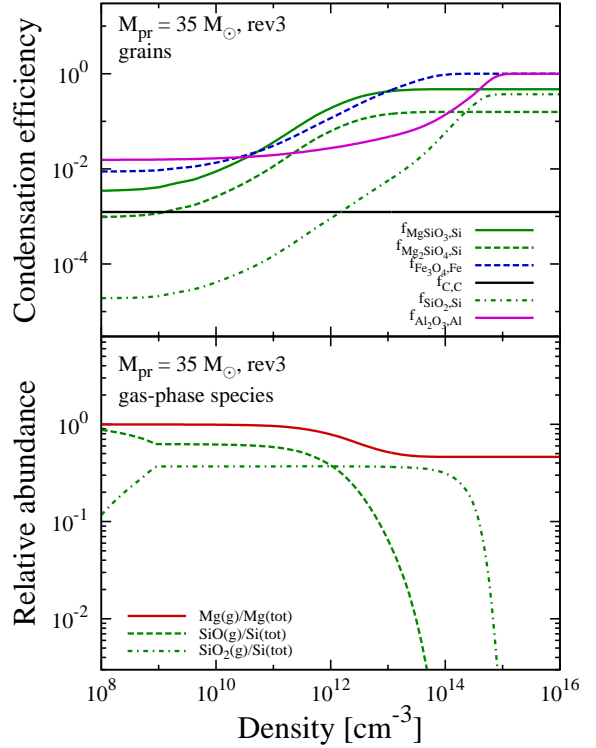}
\caption{
{\it Top panel }: the evolution of the condensation efficiency 
as a function of the cloud central density.
{\it Bottom panel }: the relative abundance of gas-phase elements.
Red solid curve represents the number fraction of gas-phase magnesium atom
relative to total Mg nuclei.
Green dashed and dotted curves depict the number fraction
of SiO and SiO$_2$ molecules relative to total Si nuclei, respectively.
}
\label{fig:nf_S0_M35_rev3}
\end{figure}

\begin{figure*}
\includegraphics[width=15cm]{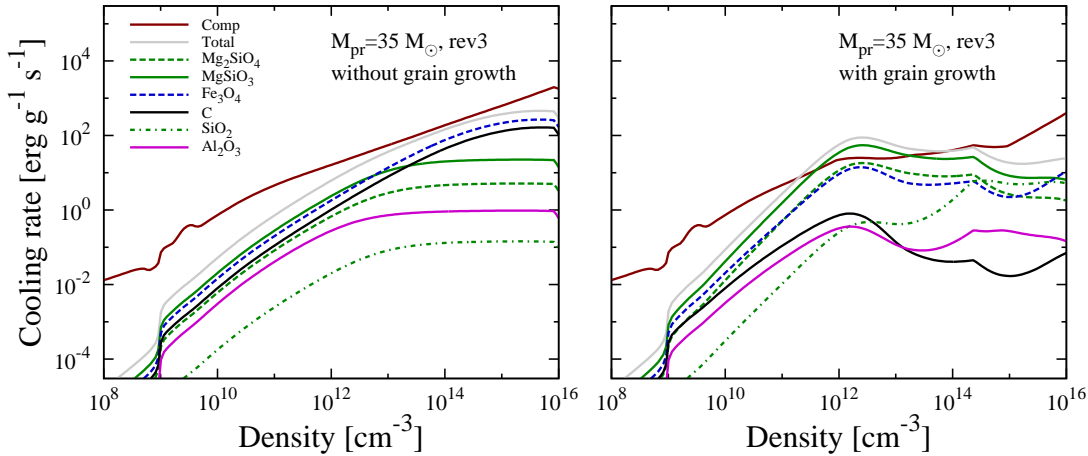}
\caption{
Compressional heating (red) and dust cooling rates as a function of gas density
for \revthree \ models with $\Mpr = 35 \ \Msun$ with (left) and without (right) grain growth.
Each line represents the contribution of a single dust species 
and the light grey solid line shows the total cooling rate. 
The abrupt change of slope at $\nH \simeq 10^{14} \ \percc$ is where to continuum 
opacity becomes effective.}
\label{fig:nLd_S0_M35_rev3}
\end{figure*}

\section{RESULTS}

In the left panels of Fig. \ref{fig:nT}, we show the predicted temperature evolution of collapsing clouds for all the
initial conditions (progenitor masses and $\fdepini$) considered in this paper.
The points mark the state where fragmentation conditions are met \citep{SchneiderOmukai10}.
Molecular cooling leads to a first temperature dip at intermediate
densities $n \approx 10^7 \ \percc$ where gravitational fragmentation is expected to 
form relatively massive fragments, with tens of solar masses. 
At higher densities, $n > 10^{11} \ \percc$, 
grain growth has a large effect on the thermal evolution of the clouds: 
all the models show that enhanced dust cooling leads to a rapid 
decrease in gas temperature and fragmentation into sub-solar mass
fragments, independently of $\fdepini$. This is different from what was 
originally found by \cite{Schneider12Caf} where 
dust cooling was unable to activate fragmentation for \revtwo \ and \revthree \ models with $\Mpr = 20 \ \Msun$,
and \revone, \revtwo, and \revthree \ models with $\Mpr = 35 \ \Msun$, which did not satisfy the condition of equation
(\ref{eq:SD_crit}).
To better clarify the impact of grain growth, in the right panel of Fig. \ref{fig:nT}
we compare the thermal evolution of the \norev \ and \revthree \ models with and without grain growth.\footnote{Our 
calculations without grain growth are the same as the calculations
performed by \citet{Schneider12Caf}.}
At densities $n < 10^{11} \ \percc$ 
the thermal tracks are still hardly affected by grain growth
because dust cooling itself is not effective in this regime.
The \norev \ tracks appear to be only moderately affected because
almost all refractory elements are already depleted (0.93 of Mg, 1.0 of Fe for $\Mpr = 20 \ \Msun$)
if they are not destroyed by the effects of
reverse shocks.
Conversely, the evolution of \revthree \ models, that suffers the strongest
dust destruction by the reverse shock and have very low initial depletion factors, 
show that grain growth leads to efficient cooling and 
fragmentation at $n > 10^{11} \ \percc$, 
where the tracks strongly deviate from the cases without grain growth.

We first discuss the results for \revthree \ model with $\Mpr = 35 \ \Msun$.
In fig.~\ref{fig:dist_S0_M35_rev3}, for each dust species we show the predicted grain size distribution functions, 
$n_i \varphi _i (r)/\nH$ in unit of ${\rm cm^{-1}}$ before 
(upper panel at $\nH = 0.1 \ \percc$) and after 
(bottom panel at $\nH = 10^{16} \ \percc$) grain growth. We can see that the radii of all grain species, 
except for carbon dust, increase. The size distribution of carbon grains is nearly unaffected because C atoms
are already depleted into CO molecules  early at densities below $\nH \sim 10^6 \ \percc$.
Since the growth rate, $(dr/dt)_i$ is independent of grain radius (equation (\ref{GG:drdt})), the net effect
is an overall shift of the size distribution function to larger radii. 
For the same model we also show in Figure \ref{fig:nf_S0_M35_rev3} the evolution of the 
condensation efficiencies and the fractional abundance of gas-phase elements as a function of the cloud central
density.
As expected, the condensation efficiencies of carbon atoms remains constant and almost equal to its initial value. 
The first grain species to grow at $\nH \sim 10^{9}$--$10^{12} \ \percc$ are silicate ($\Olivine$ and $\Pyroxene$, 
solid and dashed lines in the upper panel, respectively) grains 
whose key species is the SiO molecule. As silicate grains grow, the abundance of SiO molecules decreases until its 
abundance eventually drops to zero (see the dashed line in the bottom panel).
The growth of silica grains is controlled by SiO$_2$ molecules whose formation rate
grows with density. 
Though the abundance of SiO$_2$ molecules remains nearly
constant because the abundance of $\Silica$ grains are small 
at $\nH \sim 10^{9}$--$10^{13} \ \percc$, it eventually drops to zero 
(dot-dashed line in the bottom panel).
The condensation efficiency of silicon atoms onto silica grains grows by more than four orders of magnitudes.
Finally, although with different evolutionary properties, 
Fe and Al condensation efficiencies 
reach unity, meaning that all the available Al and Fe grains are condensed into solid grains.

We consider that
grain growth is significant when the collapse has reached the density (time) threshold, 
$\ngrowi$, at which the condensation efficiency becomes 0.5. 
From the results of calculations, we find a fitting function of
$\ngrowi$ as
\[
\ngrowi 
=
1.0\E{12} \ \percc
\left( \frac{A_j}{7.1\E{-10}} \right) ^{-2}
\left( \frac{f_{ij,0}}{0.1} \right) ^{-0.8}  \nonumber
\]
\[
\times 
\left( \frac{r_{i,0}^{{\rm grow}}}{0.01 \ \um} \right) ^{2}
\left( \frac{a_{i,0}}{1 \ {\rm \AA}} \right) ^{-6}
\left( \frac{m_{i1}}{\mH} \right),
\]
where $r_{i,0}^{{\rm grow}}$ is 
the mass-weighted average radius, $r_{i,0}^{{\rm grow}}=\left\langle r^3 \right\rangle _{i,0}^{1/3}$.
Hence, the condition for efficient grain growth depends on specific properties of the
grain species as well as on the abundance of the key species. As an example,
grains with larger $a_{i,0}$, such as silicates, grow earlier in the evolution (at lower
densities), whereas in \revthree \ model with $\Mpr=35 \ \Msun$, aluminium abundance is so small 
that the growth of $\Alumina$ grains is inefficient at all but the largest densities. 

\begin{table*}
\caption{Critical abundances
\label{tab:Crit}}
\begin{tabular}{@{}cccccccc}
\hline
$\Mpr \ (\Msun)$ & model & 
$r_{\Pyroxene}^{\rm cool} \ (\um)$ & $r_{\Magnetite}^{\rm cool} \ (\um)$ & 
$f_{\Pyroxene, \Mg}$ & $f_{\Magnetite,\Fe}$ &
${\rm [Mg/H]_{\rm cr}}$ & ${\rm [Fe/H]_{\rm cr}}$ \\
\hline 
20  & {\tt norev} &  8.39e-03 &  2.38e-02 &   0.20 &   1.00 &  -5.02 &  -4.92 \\
  &   {\tt rev1} &  1.57e-02 &  2.44e-02 &   0.35 &   0.43 &  -4.99 &  -4.53 \\
  &   {\tt rev2} &  2.31e-02 &  2.55e-02 &   0.37 &   0.19 &  -4.84 &  -4.16 \\
  &   {\tt rev3} &  3.27e-02 &  2.78e-02 &   0.40 &   0.07 &  -4.73 &  -3.70 \\
\hline \hline
 &  & 
$r_{\Pyroxene}^{\rm cool} \ (\um)$ & $r_{\Magnetite}^{\rm cool} \ (\um)$ & 
$f_{\Pyroxene, \Si}$ & $f_{\Magnetite,\Fe}$ &
${\rm [Si/H]_{\rm cr}}$ & ${\rm [Fe/H]_{\rm cr}}$ \\
\hline 
35  &  {\tt norev} &  1.45e-02 &  7.27e-03 &   0.60 &   1.00 &  -5.24 &  -5.43 \\
  &   {\tt rev1} &  1.89e-02 &  1.01e-02 &   0.44 &   0.55 &  -4.99 &  -5.03 \\
  &   {\tt rev2} &  2.51e-02 &  1.19e-02 &   0.32 &   0.28 &  -4.72 &  -4.66 \\
  &   {\tt rev3} &  3.24e-02 &  1.43e-02 &   0.19 &   0.12 &  -4.39 &  -4.21 \\
\hline
\end{tabular}
\\ \medskip
Note --- We calculate the critical abundances of Si or Mg, and Fe (seventh and
eighth columns, respectively) by using
values obtained from our calculations at $\nH = 10^{12} \ \percc$ (third to sixth columns).
\end{table*}

While the growth rate of each specific grain depends also on the initial composition of
gas-phase elements, we find that grain growth is able to activate dust cooling and 
fragmentation for all the explored initial conditions (two supernova progenitor models,
a total gas metallicity set to reproduce the observed surface metallicity of \Caf,
and $0.001 <\fdepini < 0.26$).  
In Fig. \ref{fig:nLd_S0_M35_rev3}, we show the cooling rate for \revthree \ models with 
$\Mpr = 35 \Msun$, each line representing the contribution of a single dust species.
Left panel shows the result without grain growth. 
In this case, total dust cooling does not exceed gas compressional 
heating. This means that the gas does not meet the condition for fragmentation.
With grain growth (right panel), 
dust cooling is increased and dominates over compressional heating rate.

The dominant coolants are silicate and magnetite grains. 
Even if only $\Pyroxene$ grains are present, their effect on the cooling
rate would be enough to overcome compressional heating and enable gas fragmentation.
This conclusion is independent of the specific reverse
shock model (the value of $\fdepini$) as silicates and magnetite 
grains always represent the dominant dust species.
Compressional heating is smaller for the model with grain growth because
the rate is proportional to temperature which is reduced by enhanced dust cooling.
Cooling rates increase with density 
for $10^9 \ \percc < \nH < 10^{12} \ \percc$, as a consequence of the combined
effects of grain growth and the increasing collisional rate between dust and
gas particles. Above $\nH \sim 10^{12} \ \percc$, the cooling rates start to decline
as the gas temperature approaches the dust temperature. 
For silicates and magnetite
grains this decrement is partly compensated by grain growth. At $\nH \sim 10^{14} \ \percc$, the
gas cloud starts to be optically thick to continuum opacity.

\section{DISCUSSION}

The results discussed above have been obtained under the 
assumptions of a sticking probability of
$\alpha_i = 1$ for all dust species. In addition, there
may be physical mechanisms that limit grain growth, such as
dust evaporation, and dust coagulation due to grain-grain collisions.
However, none of these affects the gas fragmentation properties,
at least in the models that we have investigated, as we explain 
in what follows.

We first consider the effect of a smaller sticking probability.
In fact, the exact value of the sticking probability for each dust species 
is unknown.
A more careful treatment might be required for grain species onto which more 
than one gas-phase element is accreted such as silicates.
Let us assume that two elements, 1 and 2, are accreted onto a grain species $i$,
with corresponding growth rates $(dr/dt)_{i1}$ and $(dr/dt)_{i2}$, respectively.
If element 1 is the key species, then by definition $(dr/dt)_{i1} < (dr/dt)_{i2}$
and the growth rate of $i$'th grains is set to be $(dr/dt)_i = (dr/dt)_{i1}$.
As long as $(dr/dt)_{i1} \ll (dr/dt)_{i2}$, 
the sticking probability would not be affected by the species which 
collides with the second most frequent collision rate.
However, when $(dr/dt)_{i1} \sim (dr/dt)_{i2}$, element 2 collides less frequently and
the sticking probability might be further reduced.
We also run models assuming that $\alpha_i = 0.1$ 
for all dust species. In this conservative limit, dust formation in the SN ejecta is also
affected and $\fdepini$ are further reduced with respect to the reference model. However,
we find grain growth to be effective and the fragmentation condition to be verified 
for all progenitor and reverse shock models.

The effects of dust coagulation in star-forming clouds with different initial metallicities
have been discussed by \citet{Hirashita09}. They show that as long as the grain velocities
are governed by thermal motions, coagulation occurs at densities 
$\nH $$> $$10^7$$(Z/Z_{\odot})^{-2}$$(T/100 \ {\rm K})^{-1} \ \percc$ but it does not have significant 
effects on the thermal evolution of clouds. In addition, grain growth increases the average
grain mass, thereby decreasing their thermal velocity which eventually drops below the 
coagulation threshold velocity. Dust evaporation may be important if the dust temperature
becomes comparable to the sublimation tempertature. To check the importance of this effect,
we have conservatively assumed that grains whose temperature
is above the sublimation temperature are immediately returned to the gas-phase.
The sublimation temperatures as a function of density are taken from \citet{Pollack94}.
We find that dust evaporation can be safely ignored because the
dust temperature always remains well below the sublimation temperature for all the models 
which we investigate here.

\section{critical initial conditions for the formation of \Caf}

The results of the present study may have important implications for the formation 
of the first low-mass stars in the Universe. 
A smaller dust-to-gas mass ratio may be 
sufficient to activate dust-induced fragmentation by the accretion of heavy elements onto grains 
during cloud collapse.
In particular, our analysis shows that the dust species that dominate cooling
are silicates and magnetite. Grain growth do not affect carbon grains, even if
these are initially present, because carbon atoms are rapidly locked into CO
molecules during the first stages of the collapse. Hence, the conditions for 
low-mass star formation may rely on the initial abundance of other refractory elements, such as 
key elements of silicates and magnetite:
Mg and Fe for our $\Mpr = 20 \ \Msun$ models, and
Si and Fe for our $\Mpr = 35 \ \Msun$ models.

A conservative lower limit to the initial abundance of refractory elements 
can be obtained as follows.
The critical condition expressed by equation (\ref{eq:SD_crit}) depends on the product
of the dust-to-gas mass ratio and the grain cross section per unit mass of dust. 
If we assume a {\it single} dominant grain species $i$, the critical condition
{\it after grain growth} can be written as,
\begin{equation}
S{\cal D} = \XH A_j f_{ij} \mu _i
\frac{\pi \left\langle r^2 \right\rangle _i}{(4\pi /3) s_i \left\langle r^3 \right\rangle _i} 
\geq S {\cal D} _{{\rm cr}},
\label{eq:crit}
\end{equation}
where $s_i$ is the material density of grains $i$.
Here, we define another characteristic size for dust cooling as
$r_{i}^{{\rm cool}} = \left\langle r^3 \right\rangle _i / \left\langle r^2 \right\rangle _i$.
From equation (\ref{eq:crit}), the critical abundance reads,
\[
A_{j,{\rm cr}} = 1.4 \times 10^{-3} \frac{4 \, s_i \, r_{i}^{{\rm cool}}}{3\XH \, f_{ij}\, \mu _i}
\left( \frac{T}{10^3 \ {\rm K}} \right) ^{-1/2}
\left( \frac{\nH}{10^{12} \ \percc} \right)^{-1/2},
\]
\noindent
for $(i,j) = (\Pyroxene , \Si \ {\rm or} \ \Mg)$ and $(\Magnetite , \Fe)$, respectively.
Table \ref{tab:Crit} shows the critical abundances that correspond to 
the values of $r_{i}^{{\rm cool}}$ and $f_{ij}$ at $\nH=10^{12} \ \percc$,
obtained for each of the supernova model that we have explored. Given the large variation 
of condensation efficiencies experienced by the key elements in the different models, we
can take as indicative critical abundances the maximum and minimum values that we find for each
refractory element:
$\rm [Mg/H] _{\rm cr} = [-5.02 : -4.73]$,
$\rm [Si/H] _{\rm cr} = [-5.24  : -4.73]$, and
$\rm [Fe/H] _{\rm cr} = [-5.43 : -3.70]$,
where we have normalized to solar abundances.\footnote{We have considered 
$A_{\rm Fe, \odot} = 3.31 \times 10^{-5}$, $A_{\rm Si,\odot} = 3.31 \times 10^{-5}$, 
and $A_{\rm Mg, \odot} = 3.47 \times 10^{-5}$ \citep{Caffau11Nat}. Also, 
the mean molecular weights per unit key elements are $\mu _{\Pyroxene} = 100$
and $\mu _{\Magnetite} = 77.3$.
The material densities are $s_{\Pyroxene} = 3.21 \ {\rm g \ cm^{-3}}$ and
$s_{\Magnetite} = 5.25 \ {\rm g \ cm^{-3}}$ \citep{Nozawa03}.} 

The initial abundances of Mg, Si, and Fe predicted by the inital supernova
progenitors are
$\rm [Mg/H] = -4.02$ and $\rm [Fe/H] = -4.55$ for the $\Mpr = 20 \ \Msun$ models and
$\rm [Si/H] = -4.39$ and $\rm [Fe/H] = -4.71$ for the $\Mpr = 35 \ \Msun$ models with $Z=4.5\E{-5} \ \Zsun$.
Hence, for all our models, the Si or Mg abundances are sufficient or marginally sufficient 
(for \revthree \ model with $\Mpr = 35 \ \Msun$) to enable fragmentation even if there were 
(hypothetically) only $\Pyroxene$ grains in the clouds. Conversely, in all models but 
the \revone \ model with $\Mpr = 35 \ \Msun$, the Fe abundance is not sufficient to enable 
fragmentation if only $\Magnetite$ grains are present in the clouds. These conclusions are
consistent with the results obtained by the full calculations shown in Fig.~1 and in the
right panel of Fig.~4.

Recently, Ji, Frebel \& Bromm (2013) have compared critical silicate
abundances for dust-cooling to chemical abundances of metal-poor stars, with [Fe/H]$< -4$.
Assuming silicon-based dust with different grain size distributions, they conclude that, out of the 9 stars
considered, 2-4 stars show silicate abundances below the critical ones, pointing
to a different formation pathway. Interestingly enough, all the stars 
that can not form through silicon-based dust satisfy the criterion for O {\sc i} and 
C {\sc ii} fine-structure line cooling \citep{Frebel07}.

We conclude that low-mass star formation induced by dust cooling may be more common than 
previously thought based on the required initial dust-to-gas mass ratio of very low-metallicity
star-forming clouds. Although the first supernovae may already provide enough dust grains,
our study --- which is motivated by the observed properties of \Caf --- 
suggests that even in the extreme conditions where a large fraction of the newly
formed dust is destroyed by the reverse shock of the supernova, 
grain growth during the collapse is able to replenish the required amount of dust
provided that a minimum abundance of refractory elements (mostly Si, Fe, and Mg) 
is initially present in the gas-phase.

\section*{acknowledgments}

We thank Simone Bianchi for his kind collaboration.
GC is supported by Research Fellowships of the Japan
Society for the Promotion of Science (JSPS) for Young Scientists.
This work is supported by World Premier International Research Center Initiative (WPI Initiative), 
MEXT, Japan and in part by Grant-in-Aid for Scientific Research 
from the JSPS Promotion of Science (23540324, 25287040, 22684004, and 23224004).
The research leading to these results has received funding from the European 
Research Council under the European Union's Seventh Framework 
Programme (FP/2007-2013) / ERC Grant Agreement n. 306476.


\end{document}